# Beam energy online measurement of BEPCII LINAC

WANG Shao-Zhe(王少哲)[1,2] LIU Rong(刘熔)[1] CHI Yun-Long(池云龙)[1]

[1]Laboratory of Particle Acceleration Physics & Technology, Institute of High Energy Physics, Chinese Academy of Sciences, Beijing 100049, China

[2]University of Chinese Academy of Science, Beijing 100049, China

**Abstract:** This paper describes beam energy online measurement of BEPCII linac, presents the calculation formula and some of the results. The method mentioned here measures the beam energy by acquiring beam positions in the horizontal direction with three beam position monitors (BPM) eliminating the effect of orbit fluctuation, which is much better than the one using the single BPM. The error analysis indicates that this online measurement has further potential usage such as a part of beam energy feedback system. The reliability of this method is also discussed and demonstrated in the end of this paper.

**Key words:** BEPCII linac, Beam energy, Online measurement

**PACS:** 29.20.Ej; 29.27.Eg; 29.27.Fh

## 1 Introduction

After Beijing Electron Positron Collider (BEPC) being upgraded into BEPCII, its performance has been greatly improved. The beam energy supplied by the linear accelerator (linac) for its storage ring is increased from the range of 1.1 to 1.55GeV to the range of 1.89 to 2.5GeV [1]. Especially, the phase control system and the sub-harmonic system added into the linac play an important role during the operation of the injector and make it more stable. Under the cooperation of different systems, the injection rate of positron increases from 5mA/min to 50mA/min, and the injection rate of electron also has an obvious improvement [1]. However, there are not so many systems we can depend on to measure the beam energy in the part of linac except two named AM2 and AM3, and neither of them works during the injection. This paper is aimed to solve the problem that how to do the real time measurement of the beam energy for BII linac. According to the actual situation of BII, we use a group of position values obtained by three BPMs to calculate the difference between the real energy and the nominal one. Because of its reduction on the influence of orbit fluctuation, the method is advanced compared with the one using the single BPM in the place of large dispersion.

## 2 Beam energy online measurement

There are a series of bending magnets in the transport line between the end of BII linac and the storage ring. They make the dispersion function not equal to zero on the orbit of the beam. If a BPM is put in a place of large dispersion, the distance between the position of beam and the center trajectory in x direction it obtained can be used to calculate the difference between the real energy and the nominal one. Equation (1) [2] shows the relationship between the distance and the energy difference, in which Δx means the distance between the position of beam and the position of the central trajectory, η is the value of dispersion function, E is the nominal energy and ΔE is the energy difference.

$$\Delta x = \eta \frac{\Delta E}{E} \qquad (1)$$

The equation above is correct only without the effect of orbit fluctuation. Actually, it is impossible. The Δx obtained by the BPM contains two parts, one is the position deviation in the place of large dispersion caused by the energy difference or momentum difference, the other is the beam orbit fluctuation [3]. The latter has nothing to do with the beam energy. To make the energy difference more accurate, one must eliminate the effect of orbit fluctuation. Because of this we need at last tree BPMs.

In order to reduce the error, two of the three BPMs should be located in the place behind the last accelerating tube at the end of the linac and before the first BPM in big dispersion. Fortunately, there are four BPMs whose resolution is high enough meeting all of the conditions above in the transport line of BII, seen in Fig. 1. They are TCBPM1 and TCBPM2 in the common line, TEBPM1 in the electron line and TPBPM1 in the positron line. Each kind of beam uses three BPMs for energy measurement and TCBPM1 and TCBPM2 are public for both kind.

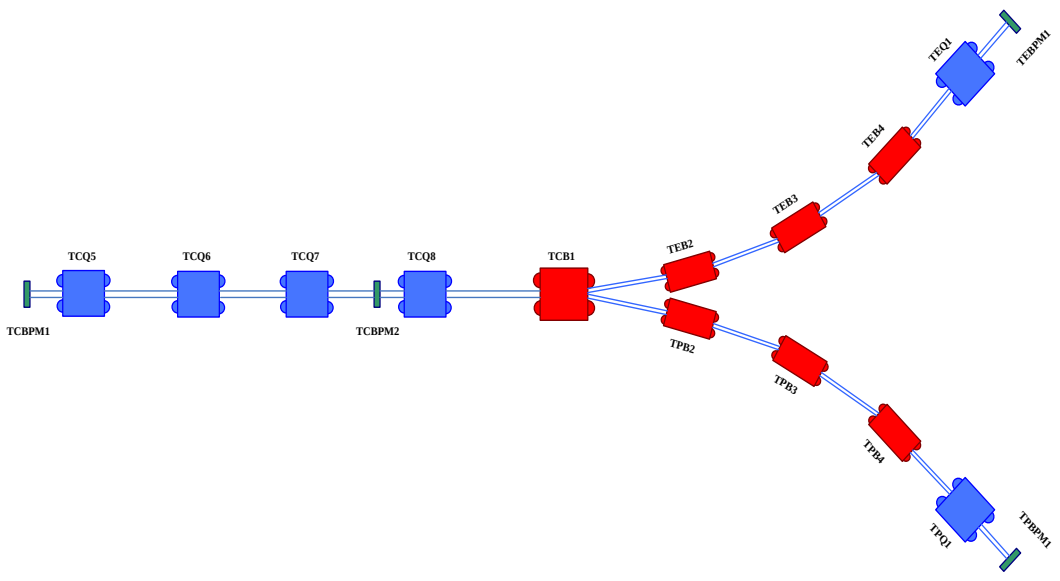

Fig. 1. The layout of BPMs at the end of BII linac.

Taking electron as an example, we first get the distance between the center of beam and the ideal orbit via TCBPM1 and TCBPM2, then calculate the momentum bias by using the transfer matrix and TEBPM1 measurement value. And the energy bias can be obtained by equation (2) [4], in which ΔP is the momentum bias, P is the design momentum and β is the ratio of the velocity and speed of light.

$$\frac{\Delta E}{E} = \beta^2 \frac{\Delta P}{P} \qquad (2)$$

It is necessary to state some conditions and make some assumptions. (1) The BII linac accelerates electrons and positrons. When the beam energy reaches 1.89GeV, the Lorenz coefficient γ will be bigger than 3600, which indicates that the space charge effect can be neglected [5]. (2) Suppose that the bunch has a two-dimensional Gauss distribution in the x-x' phase space [6]. (3) Don't consider the coupling between the longitudinal direction and the transverse direction in the straight part [7]. (4) Consider the influence to the horizontal direction caused by the momentum bias in the place where the value of the dispersion function is not equal to zero.

If a bunch of electrons come across TCBPM1, TCBPM2 and TEBPM1 orderly, the distance values obtained by the BPMs are x1，x2 and x3 respectively, and correspondingly the angles are x1'，x2' and x3', we can get equations (3) and (4) with the help of transmission matrix. In the equations, M1x is the transmission matrix of the phase space in the horizontal direction from TCBPM1 to TCBPM2, and M2x is the transmission matrix of the phase space in the horizontal direction from TCBPM2 to TEBPM1.

$$\begin{bmatrix} x_2 \\ x_2' \end{bmatrix} = M_{1x} \begin{bmatrix} x_1 \\ x_1' \end{bmatrix} \qquad (3)$$

$$\begin{bmatrix} x_3 \\ x_3' \\ \frac{\Delta P}{P} \end{bmatrix} = M_{2x} \begin{bmatrix} x_2 \\ x_2' \\ \frac{\Delta P}{P} \end{bmatrix} \qquad (4)$$

$$(M_{1x} = \begin{bmatrix} M_{1x}(11) & M_{1x}(12) \\ M_{1x}(21) & M_{1x}(22) \end{bmatrix},$$

$$M_{2x} = \begin{bmatrix} M_{2x}(11) & M_{2x}(12) & M_{2x}(13) \\ M_{2x}(21) & M_{2x}(22) & M_{2x}(23) \\ 0 & 0 & 1 \end{bmatrix} )$$

While *γ>>1*, and $\beta = \sqrt{1 - 1/\gamma^2}$, so *β≈1*, then the equation (2) can approximately be Δ*E/E*≈Δ*P/P*. From (5), we can get (6), which is the expression of energy difference.

$$\frac{\Delta E}{E} = k_1 x_1 + k_2 x_2 + k_3 x_3 \qquad (6)$$

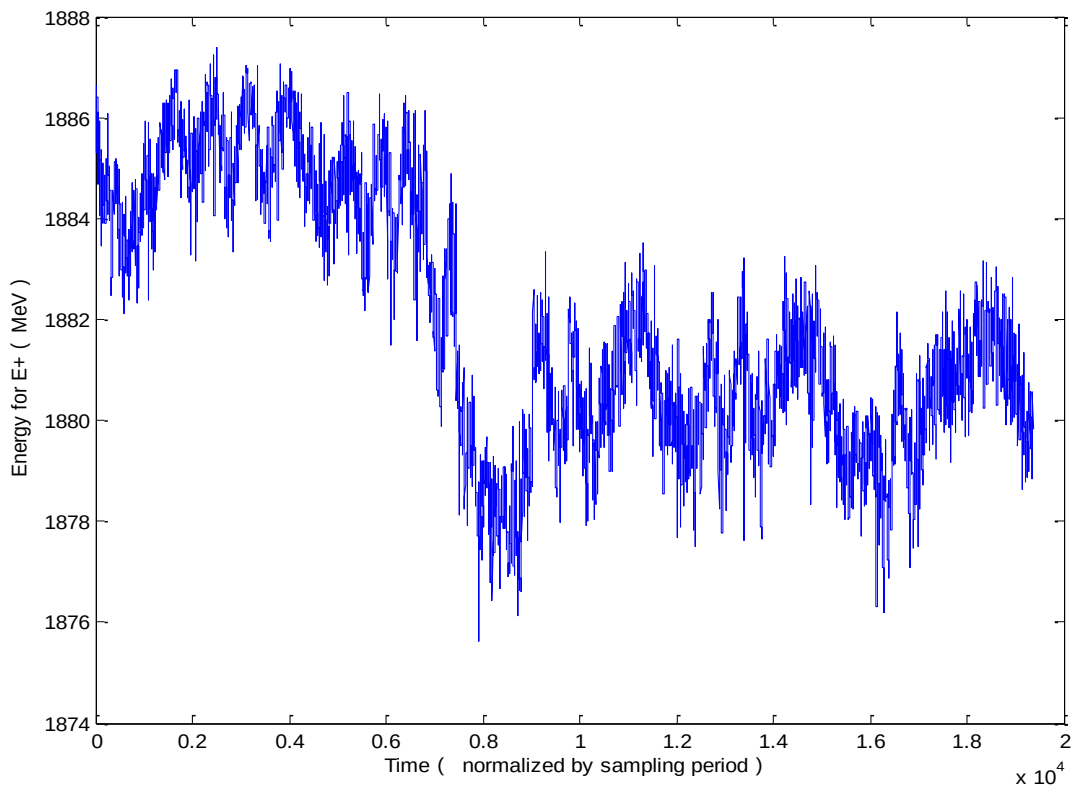

Fig. 2. The result of the measurement of the beam energy for BII positron injection (19:55 3/9/2015).

According to the situation of BII, the two order matrix $M_{1x}$ is the product of transmission matrix of phase space in the horizontal direction for different elements, named the drift tube, the defocusing quadrupole lens (just for electrons, considering positrons, it will be on the contrary), the drift tube, the focusing quadrupole, the drift tube, the defocusing quadrupole lens and the drift tube orderly. While the three order matrix $M_{2x}$ is also the product of a series of transmission matrix of the elements. What the difference is there's a new kind of element called the bending magnet in it. Depending on the parameters of BII transport line, these two matrixes can be calculated. Then under the help of EPICS systems for the related values of BPMs and the magnet currents, we can get the beam energy during the injection. Fig. 2 shows the result of the measurement of beam energy using this method. It's a positron injection at 19:55, Mar. 9th, 2015.

## 3 Error analysis

There are some sources that may cause the error: (1) the installation error of BPMs, which can be regarded as the difference of the measuring center and the orbit center in the horizontal direction, and the difference of the real distance and the nominal distance of BPM and magnet in the longitudinal direction; (2) the resolution of BPMs; (3) the installation error of dipoles and quadrupoles, which includes 6 dimensions such as displacement in the horizontal and longitudinal directions, rotation in the horizontal direction and rotation in longitudinal direction; (4) the influence of the current stability when the magnets are working; (5) the error of magnetic field gradient caused by interpolation.

The coupling effect of the factors above is very weak, so that they can be considered independent of each other. The total effect can be expressed by equation (7).

$$\delta = \sqrt{\sum_i \delta_i^2} \qquad (7)$$

The errors about magnets make the coefficients in the functions different to their nominal values. However, calculating the effect directly is very difficult. Because the number of coefficients is large and there's coupling effect between the upstream element and the downstream one. In order to estimate the error, we suppose that the coefficients in the functions are correct, what the errors cause is to make the beam positions and the beam angles we obtained inaccurate.

The function of the energy difference can be written into a first order linear one like equation (8), in which $a_1$, $a_2$, $a_3$ are the coefficients before $x_1$, $x_2$, $x_3$ respectively. They are related to the parameters of the elements ($a_1$ equals $k_1E$, $a_2$ equals $k_2E$, $a_3$ equals $k_3E$).

$$\Delta E = a_1x_1 + a_2x_2 + a_3x_3 \qquad (8)$$

Considering the relationship between $x_1$, $x_2$ and $x_3$, the error of $\Delta E$ delivered from the error of $x_1$, $x_2$ and $x_3$ is

$$\delta_{\Delta E}^2 = a_1^2\delta_{x1}^2 + a_2^2\delta_{x2}^2 + a_3^2\delta_{x3}^2 + 2a_3a_1\delta_{x3}\delta_{x1} + 2a_2a_1\delta_{x2}\delta_{x1} + 2a_3a_2\delta_{x3}\delta_{x2} \qquad (9)$$

For the value of x1 measured by TCBPM1, its error $\delta_{x1}$ has something to do with the installation error of BPM and the resolution, seen in (10).

$$\delta_{x1} = \sqrt{\delta_{T1}^2 + \delta_{R1}^2} \qquad (10)$$

For the values of $x_1$ and $x_2$ measured by TCBPM2 and TE/TPBPM1, their errors are related not only to the installation error of BPM $\delta_T$ and the resolution $\delta_R$, but also to the installation error of dipoles $\delta_{DI}$ and quadrupoles $\delta_{QI}$, the current stability $\delta_{CS}$ and the errors of interpolation $\delta_{It}$. The function is (11).

$$\delta_{x2/3} = \sqrt{\delta_{T2/3}^2 + \delta_{R2/3}^2 + \sum\delta_{QI2/3}^2 + \sum\delta_{CS2/3}^2 + \sum\delta_{DI2/3}^2 + \sum\delta_{It2/3}^2} \qquad (11)$$

Take the defocusing quadrupole lens into consideration, its installation error of the horizontal direction $\delta_{DIx}$ will make the beam position has a bias in the x direction shown in (12), in which $K$ means the magnetic field gradient of the quardrupole and $l$ means the length of the lens [8].

$$\Delta X_{Dx} = \delta_{DIx}\begin{bmatrix}1 - cosh(Kl)\\ -Ksinh(Kl)\end{bmatrix} \qquad (12)$$

What's more, the bias of the beam will be accumulated while the beam is going through the elements. At last, the position seen by the BPM will be equation (13).

$$\Delta X_{dbx} = M_{db}\Delta X_{Dx} \qquad (13)$$

In the equation, $M_{db}$ is the transfer matrix of the horizontal direction from the exit of the lens to the BPM.

Similarly, the bias caused by the installation error of the longitudinal direction $\delta_{DIz}$ is shown in equation (14), while $x_0$ is the horizontal position of the beam in the entrance of the lens and $x_0'$ is the horizontal angle of the beam in the same place.

$$\Delta X_{dbz} = M_{db}\delta_{DIz}\begin{bmatrix}-K\sinh(\mathrm{Kl})x_0\\ Ksinh(Kl)x_0'\end{bmatrix} \qquad (14)$$

The rotation error with the longitudinal axis of the lens $\delta_{DIzr}$ and the rotation error with the horizontal axis of the lens $\delta_{DIxr}$ make the bias in the equation (15) and (16), in which $y_0$ is the longitudinal position of the beam in the entrance of the lens and $y_0'$ is the longitudinal angle of the beam in the same place. However, the bias caused by the installation error of the vertical direction and the rotation error with the vertical axis is so small that can be ignored.

$$\Delta X_{dbzr} = M_{db}\delta_{DIzr}\begin{bmatrix} -2(1-\cosh(\mathrm{Kl}))y_0 \\ 2K sinh(Kl)y_0' \end{bmatrix} \quad (15)$$

$$\Delta X_{dbxr} = M_{db}\delta_{DIxr}\begin{bmatrix} -\frac{l}{2}(1+\cosh(Kl) - \frac{2\sinh(\mathrm{Kl})}{Kl} \\ -(1-\cosh(Kl) + \frac{Kl}{2} sinh(Kl)) \end{bmatrix}$$

(16)

Accordingly, the biases caused by the installation errors of the focusing quadrupoles $\delta_{FIx}$, $\delta_{FIz}$, $\delta_{FIzr}$ and $\delta_{FIxr}$ is shown from the equation (17) to (20).

$$\Delta X_{fbx} = M_{fb}\delta_{FIx}\begin{bmatrix} 1-cos(\mathrm{Kl}) \\ K sin(Kl) \end{bmatrix} \quad (17)$$

$$\Delta X_{fbz} = M_{fb}\delta_{FIz}\begin{bmatrix} -K\sin(\mathrm{Kl})x_0 \\ K sin(Kl)x_0' \end{bmatrix} \quad (18)$$

$$\Delta X_{fbzr} = M_{fb}\delta_{FIzr}\begin{bmatrix} -2(1-cos(\mathrm{Kl}))y_0 \\ -2K sin(Kl)y_0' \end{bmatrix} \quad (19)$$

$$\Delta X_{fbxr} = M_{fb}\delta_{FIxr}\begin{bmatrix} -\frac{l}{2}(1+\cos(Kl)) - \frac{2\sin(\mathrm{Kl})}{Kl} \\ -(1-\cos(Kl)) - \frac{Kl}{2} sin(Kl) \end{bmatrix}$$

(20)

The installation errors of the bending magnets, such as position error of horizontal direction $\delta_{BIx}$, position error of longitudinal direction $\delta_{BIz}$, rotation error with the z axis $\delta_{BIzr}$, rotation error with the x axis $\delta_{BIxr}$ and rotation error with the y axis $\delta_{BIyr}$, can also make the bias of the beam, which are shown from equation (21) to (25). However, the position error of the vertical direction does not affect the bias in the horizontal direction, and can be ignored. In these equations, n is the gradient index of the bending magnet. In the uniform magnetic field, n is 0. Considering that the beam is near the center of the magnet, where the field can be seen as uniform field, n is 0 here. And $K_x$ refers to $1/\rho$, which is the reciprocal of the bending radius.

$$\Delta X_{bbx} = M_{bb}\delta_{BIx}\begin{bmatrix} 1-cos(\mathrm{K}_x\mathrm{l}) \\ \mathrm{K}_x sin(\mathrm{K}_x l) \end{bmatrix} \quad (21)$$

$$\Delta X_{bbz} = M_{bb}\delta_{BIz}\begin{bmatrix} \frac{l}{\rho} - K_x sin(\mathrm{K}_x\mathrm{l})x_0 \\ \mathrm{K}_x sin(\mathrm{K}_x l)x_0' \end{bmatrix} \quad (22)$$

$$\Delta X_{bbzr} = M_{bb}\delta_{BIzr}\begin{bmatrix} -\frac{1-2n}{\rho^2 K_x^2}(1-cos(\mathrm{K}_x\mathrm{l}))y_0 \\ -\frac{1-2n}{\rho^2 K_x} sin(\mathrm{K}_x l)y_0 \end{bmatrix} (23)$$

$$\Delta X_{bbxr} = M_{bb}\delta_{BIxr}\begin{bmatrix} \frac{1}{\rho K_x^2}(1-cos(\mathrm{K}_x\mathrm{l}))y_0' \\ \frac{1}{\rho K_x} sin(\mathrm{K}_x l)y_0' \end{bmatrix} \quad (24)$$

$$\Delta X_{bbyr} = M_{bb}\delta_{BIyr} \cdot$$

$$\begin{bmatrix} -\frac{l}{2}\left(1+cos(\mathrm{K}_x\mathrm{l}) - \frac{2\sin(\mathrm{K}_x\mathrm{l})}{\mathrm{K}_x\mathrm{l}}\right) - \frac{1}{\rho K_x}\sin(\mathrm{K}_x l)\, x_0 \\ -\left(1-cos(\mathrm{K}_x\mathrm{l}) - \frac{\mathrm{K}_x\mathrm{l}\sin(\mathrm{K}_x\mathrm{l})}{2}\right) + \frac{1}{\rho K_x}\sin(\mathrm{K}_x l)\, x_0' \end{bmatrix}$$

(25)

The position bias of the beam related to the instability of the magnetic field is mainly caused by the jitter of the field current. So the value of stability equals the fluctuation of the current approximately. The biases caused by the fluctuation of the magnetic field $\delta_B$, and the fluctuation of the magnetic field gradient of defocusing and focusing quadrupole $\delta_{DG}$ and $\delta_{FG}$ are shown in the following equations.

$$\Delta X_{bbs} = M_{bb}\delta_B\begin{bmatrix} \frac{1}{\rho K_x}(1-cos(\mathrm{K}_x\mathrm{l})) \\ \frac{1}{\rho K_x} sin(\mathrm{K}_x l) \end{bmatrix} \quad (26)$$

$$\Delta X_{dbs} = M_{db}\delta_{DG}\begin{bmatrix} (1-cosh(\mathrm{Kl}))x_0 \\ K sinh(\mathrm{K}l)x_0 \end{bmatrix} \quad (27)$$

$$\Delta X_{fbs} = M_{fb}\delta_{FG}\begin{bmatrix} -(1-cos(\mathrm{Kl}))x_0 \\ -K sin(\mathrm{K}l)x_0 \end{bmatrix} \quad (28)$$

Observe equations from (13) to (28), and we may find that some kinds of bias depend on the position of the beam ($x_0$, $x_0'$, $y_0$, $y_0'$) in phase space, but others are not. Those position biases that have nothing to do with the beam positions can be corrected, because their values are relatively stable. While those related to the position of the beam cannot be corrected for the reason that their values will change when the beam position changes. So it is necessary to make a classification for the errors. Those who are independent on beam phase space and can be corrected, are called the first kind of error; while others that could not be corrected are called the second kind of error. Of particular note is that though the error caused by the instability of the magnet filed has nothing to do with the beam position, it cannot be corrected either for its randomness. It should be regarded as the second kind of error.

According to the situation of BII transmission line, each error datum together with its style is shown in *Table 1* [9]. Because it is so hard for us to acquire the real number of the error and is no need for just estimation, we substitute the maximum of the tolerance into the function. Considering the terms that related to the phase space of the beam, for the same reason, we substitute the values which are big enough, that means $|x_0|=|y_0|=5mm$ and $|x_0'|=|y_0'|=5$. The results are shown in *Table 2*. In the estimation, we choose 2.5GeV as the designed energy.

From the results in *Table 2*, though this measurement doesn't have much advantage in the term of absolute precision, it still has the feature that makes us excited. After dividing the total error into two kinds, we find that the main of the measurement error is devoted by the first kind error, which can be reduced or even eliminated by calibration. While the second kind error, whose effect is hard to be removed, is rather small. This feature indicates that the on-line method has a huge potential, especially in the aspect of control, for example it will be greatly useful in beam energy feedback systems.

Table 1. The error data.

| **The installation tolerance of the quadrupole lens** | | | | | |
|---|---|---|---|---|---|
| **Name** | **Type** | **Value** | **Name** | **Type** | **Value** |
| $\lvert\Delta x\rvert$ | First kind | 0.2mm | $\lvert\Delta\theta_x\rvert$ | First kind | 1mrad |
| $\lvert\Delta y\rvert$ | First kind | 0.2mm | $\lvert\Delta\theta_y\rvert$ | First kind | 1mrad |
| $\lvert\Delta z\rvert$ | Second kind | 1.0mm | $\lvert\Delta\theta_z\rvert$ | Second kind | 2mrad |
| **The installation tolerance of the bending magnet (from B-1 to B-17)** | | | | | |
| **Name** | **Type** | **Value** | **Name** | **Type** | **Value** |
| $\lvert\Delta x\rvert$ | First kind | 0.5mm | $\lvert\Delta\theta_x\rvert$ | Second kind | 0.5mrad |
| $\lvert\Delta y\rvert$ | First kind | 0.5mm | $\lvert\Delta\theta_y\rvert$ | First + Second kind | 0.5mrad |
| $\lvert\Delta z\rvert$ | First + Second kind | 1.0mm | $\lvert\Delta\theta_z\rvert$ | Second kind | 0.6mrad |
| **Other data** | | | | | |
| **Name** | | **Type** | | **Value** | |
| Stability of the quadrupole lens | | Second kind | | 0.001 | |
| Stability of the bending magnet | | Second kind | | 3*0.0001 | |
| BPM resolution | | Second kind | | 0.05mm | |
| BPM installation tolerance | | First kind | | 1mm | |
| Field gradient interpolation error | | Second kind | | 0.001 | |

Table 2. The result of the error estimation.

| **The sum of the first kind and the second kind** | | **The error of second kind** | |
|---|---|---|---|
| $\delta_{x1}$ | 1mm | $\delta_{x1}$ | 0mm |
| $\delta_{x2}$ | 1.14mm | $\delta_{x2}$ | 0.03mm |
| $\delta_{x3}$ | 1.20mm | $\delta_{x3}$ | 0.03mm |
| $\delta_{\Delta E}$ | 3.30MeV | $\delta_{\Delta E}$ | 0.06MeV |

## 4 Verification of online measurement

From different aspects, we observed the values we obtained and the results we calculated to make sure that the measurement is reliable.

Firstly, we established a database to record the data from the EPICS system. Fig. 3 shows the record of the phase of K16, the position values of TCBPM1, TCBPM2, TEBPM1 and TPBPM1 during the positron injection. It can be seen clearly that in the period of injection (from about 20:28 to 20:43), the values of TCBPM1 and TCBPM2 change slightly, while the values of TPBPM1 have a violent oscillation and the values of TEBPM1 are noise only, which fits the expectation well. The reason is that TCBPM1 and TCBPM2 are in the straight part, so the bias of the beam position depends on the orbit fluctuation only. When the magnets of the linac works well, the range of the bias cannot be large. However, the values read by TPBPM1 contains not only the orbit fluctuation but also the effect of the beam energy, which is dominant, so the changes will be large. All of the things above indicates that the performance of the BPMs is reliable.

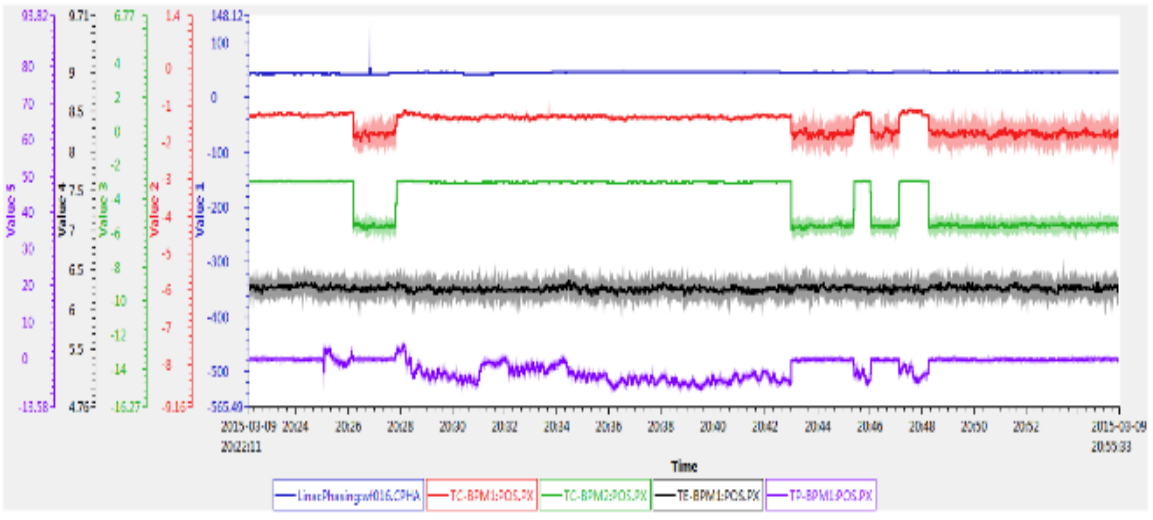

Fig. 3. The phase of K16 and the values of BPMs recorded during the positron injection of BII.

There's an offline beam energy measurement system called AM3 in BII linac. The beam is deviated out by an analysis magnet at the end of the accelerator, and goes through a slit and a BPM. After calibration, the field current of the magnet corresponds to the beam energy. And it can judge that if the beam goes through the center of the magnet by observing the values of the BPM. In order to make the comparison, we first record the values of the BPMs and the current of the magnets during a period of injection to calculate the energy. Followed by this, we turn on the AM3 system without changing other conditions. By adjusting the field current of analysis magnet, we make the beam go through the center, and record the current. According to the corresponding relationship, we may get the beam energy, compared with the online result in Fig. 4. The red points refer to the online results and the blue ones refer to the offline results. From the picture, we find that the online results are in the range of the offline results. They are nearly the same in the term of average value. However, considering the effect, the error of the offline method is bigger because of the influence of the fluctuation of the field current. So the online measurement is superior to the AM3 measurement.

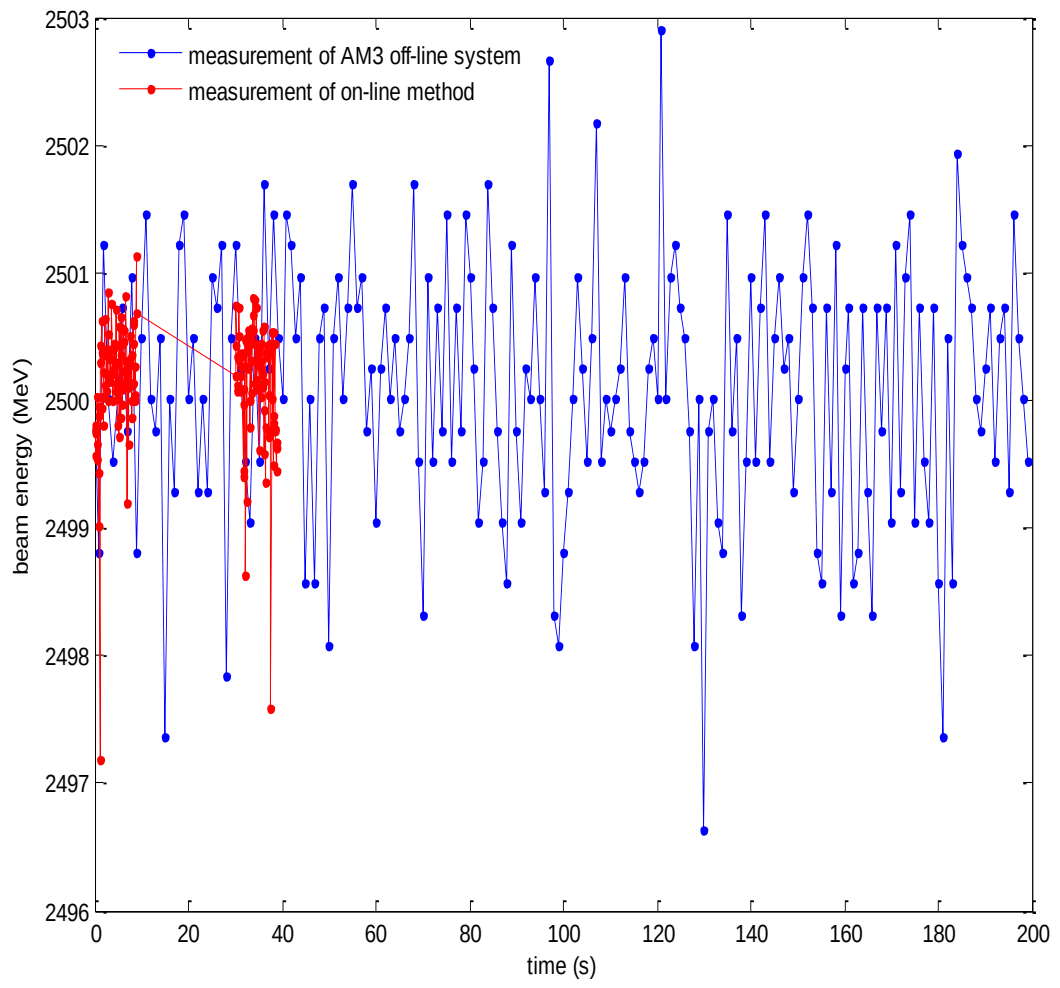

Fig. 4. The comparison of AM3 system and online measurement.

Another intuitive way is to observe the change of the beam energy with the change of the phase of klystron No.16. For the reasons of stability and steerability, the klystrons of BII linac all works in a saturated state, which means the amplitude of the output stays almost the same. In this situation, the energy gain of the beam depends on the cosine of the klystron's phase. Fortunately, one of the twenty klystrons in BII linac named K16 is usually used by the operators to adjust the phase to control the injection energy, while others works in the mode of closed loop and their phases couldn't change sharply. So there is reason to believe the beam energy is mainly controlled by K16's phase. Fig. 5 shows the curve of the cosine of K16's phase (down) and the curve of the beam energy difference gotten by online measurement (up). It's a period of positron injection from 20:28 to 20:42 on Mar.9th, 2015 as shown in Fig. 3. In Fig. 5, there is a close correlation between the beam energy difference and the cosine of the phase, which proves the effectiveness of the measurement again.

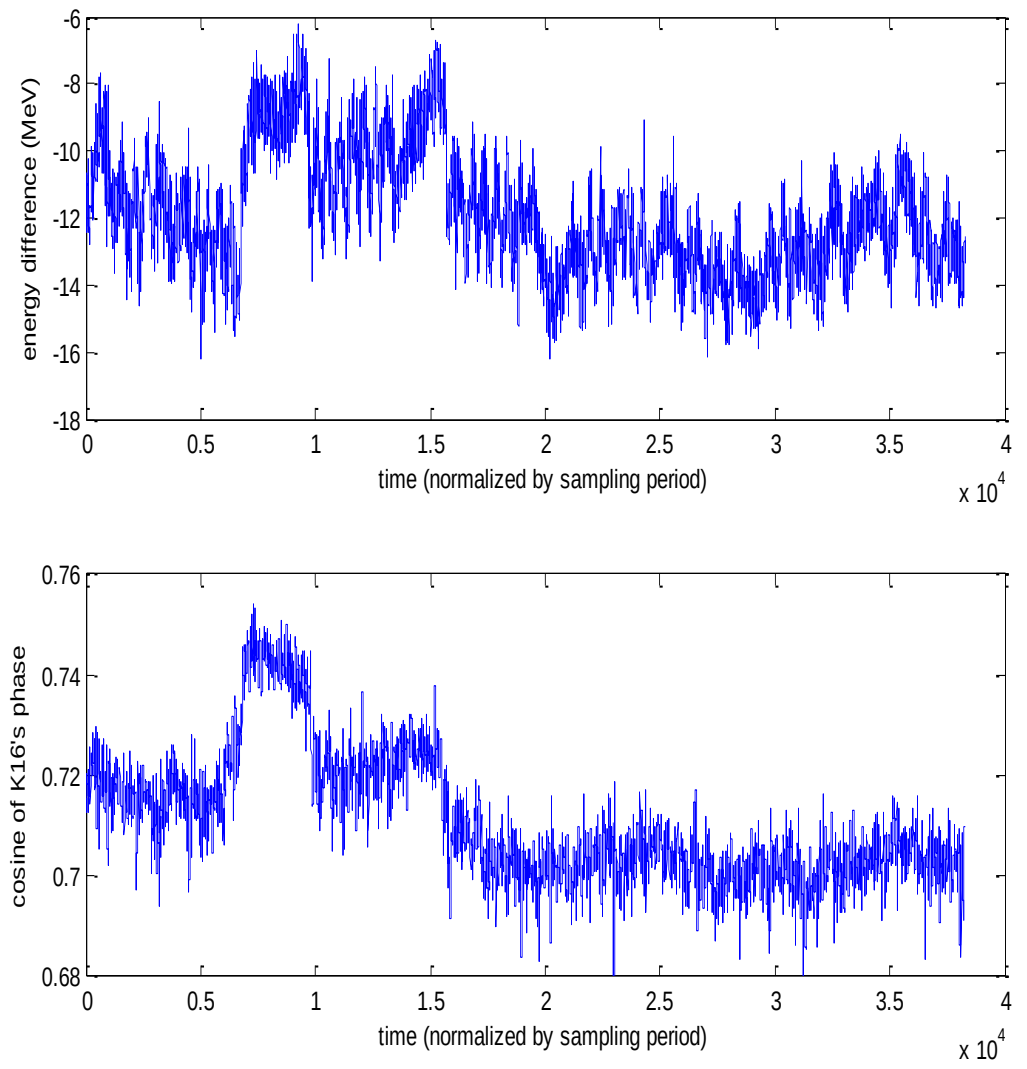

Fig. 5. The curve of the cosine of K16's phase (down) and the curve of beam energy difference obtained by on-line method (up).

## 5 Conclusion

In this beam energy online measurement method, we get the data by using three BPMs on the transfer line of BEPCII, and calculate the beam energy without some of effect from the orbit fluctuation. This real time measurement can be done during beam injection and the result shows that it is more accurate than the existing beam energy analysis system. For some point of view, this measurement is very reliable. This paper shows that we can reduce even eliminate the first kind error by doing some calibration, only leaving small amount of the second kind error. The method as a part of beam energy feedback system of BEPCII linac, will be used in the short future.

The method mentioned in this paper will be used in the development of beam energy feedback system for BEPCII linac, which is also beneficial for newly build machine such as HEPS, CEPC or other large accelerators.